\documentstyle[12pt]{article}

\begin{document}

\title{Is supernovae data in favour of isotropic cosmologies?}
\author{Mariusz P. D\c abrowski 
\footnote{email:mpdabfz@uoo.univ.szczecin.pl} \\
Institute of Physics, University of Szczecin, \\
Wielkopolska 15, 70-451 Szczecin, Poland.}

\maketitle

\begin{abstract}
Most of the observational claims in cosmology are based on the
assumption that the universe is isotropic and homogeneous so they essentially
test different types of Friedmann models. This also refers to recent
observations of supernovae Ia, which, within the framework of Friedmann
cosmologies give strong support to negative pressure matter and also weaken 
the age conflict. In this essay we drop the assumption of homogeneity, though
temporarily leaving the assumption of isotropy with respect to one point, and
show that supernovae data can be consistent with a model of the universe with 
inhomogeneous pressure known as the Stephani model. 
Being consistent with supernovae data we are able to get the age of the universe 
in this model to be about 3.8 Gyr more than in its Friedmann counterpart. 
\end{abstract}

\vspace{2.cm}
\begin{center}
{\it An essay presented for Gravity Research Foundation Essay Competition '99.}
\end{center}

\newpage

The standard isotropic cosmological models have intensively been studied
as the models of the large-scale structure of the universe. One of the main
reasons is their mathematical simplicity expressed in terms of
the Cosmological Principle. There is of course some `evidence'  
for these models to be the right ones from many different astronomical tests
and especially from low-redshift linear Hubble expansion law (e.g.
\cite{strauss}). However, the situation is not so clear for large-redshift
objects since the generalized Hubble law -- the redshift-magnitude relation -- 
becomes nonlinear and the effects of spatial curvature of the universe are
important. In the past the main problem was that the luminosity function of 
prospective `standard candles' (whose absolute magnitude is presumably 
known) was poorly known for most of them at redshifts $z \approx 1$. This
is not the case 
for supernovae type Ia (SnIa) and these objects have recently been used to 
determine the curvature and consequently the matter content of the universe
\cite{42,lambda}. The results of these investigations give strong support to
Friedmann models with {\it negative pressure} matter such as the cosmological
constant, domain walls or cosmic strings \cite{vil,negative}. It is a very
strong claim, since, despite a very long story of the cosmological
constant, \cite{einstein} and a relatively long story of topological defects
\cite{defects}, people hardly believed in their large contribution to the
total energy density of matter in the universe at the present epoch of
the evolution. 

In this essay we try to make an alternative proposal for the explanation of
supernovae data and suggest an inhomogeneous model of the universe which
belongs to the class of models known as
the Stephani universes \cite{step,mar1}. We basically try to fit this model to SnIa 
data as given in \cite{42}. Our model is described by the following
metric tensor \cite{mar2}
\begin{equation}
ds^2 = - \frac{c^2}{V^2}d\tau^2 + \frac{R^2}{V^2} \left[ dr^2 + r^2 \left(
d\theta^2 + \sin^2{\theta} d\varphi^2 \right) \right]  ,
\end{equation}
with 
\begin{eqnarray}
R(\tau) & = & a\tau^2 + \tau ,\\
V(\tau,r) & = & 1 - \frac{a}{c^2} \left(a\tau^2 + \tau\right)r^2 \\
k(\tau) & = & - 4 \frac{a^2}{c^2} R(\tau) ,
\end{eqnarray}
where $\tau$ is the cosmic time coordinate and $r$ is the radial coordinate.
In Eq. (1) $R(\tau)$ is the generalized scale factor and $k(\tau)$ 
is the time-dependent spatial curvature index, so the spatial curvature
of the universe may change during the evolution which is impossible in
Friedmann models. The constant $c$ is the velocity of light and the 
parameter $a$ is measured in ${\rm km}^2 \,{\rm s}^{-2}\,{\rm Mpc}^{-1}$. 
The physical meaning of $a$ is that it measures non-uniformity of 
pressure (acceleration) of the model (see \cite{mar2}). One is able to obtain
the flat Friedmann model from (1) is one takes the limit $a \to 0$. 

In the model described by the metric (1) the energy density $\varrho$ depends
on the cosmic time, similarly as in Friedmann models, but the pressure, $p$, is
the function of both the time and radial coordinates \cite{mar3}. This justifies its name as
`inhomogeneous pressure universe'. It is spherically symmetric and it can
certainly be used as the first step towards the observational 
verification of inhomogeneous cosmologies. A general class of 
Stephani models is really inhomogeneous which means there are no Killing
vectors in spacetime. The assumption of spherical symmetry is of course 
equivalent to dropping the assumption of the Cosmological Principle, 
provided we put an observer outside of the center of symmetry. In this 
essay we do not consider off-centre observers although the suitable
relations are known \cite{mar2}. Despite that, we can have an important
effect on the observational relations at the center, because the light
reaching the observer there was emitted from the off-center galaxies.

The spherically symmetric Stephani model is the model of concentric pressure
spheres (pressure varies from sphere to sphere) and it can be put in some
opposition to the Tolman model \cite{tolman,bonnor} which is the model 
of concentric density spheres (energy density varies from sphere 
to sphere). Both models do not necessarily have to be used as models of the
global geometry of the universe, but can also be applied to model local
inhomogeneity (or void) in the universe. Kinematically, both models expand 
and in the Tolman model there is shear while in the Stephani model there is 
acceleration. Acceleration is the result of the combined effect of 
gravitational and inertial forces on the fluid which are
unable to be separated and appears due to the spatial pressure gradient 
on the concentric spheres -- the particles are accelerated in the direction from
high-pressure regions to low-pressure regions. 

We consider our investigations of Stephani models in the context of supernovae
data as an important step towards understanding the large-scale
structure of the universe because very few, if any, attempts to compare
inhomogeneous models of the universe (see \cite{kras}) with astronomical data
have been done so far. It was done for the first time by using a preliminary 
SnIa data \cite{7} in \cite{mar3} and in this essay we try to give new 
insight into the problem using large sample data given in \cite{42}. 

The parameter space of Friedmann models contains of three parameters: the
Hubble constant $H_0$, the deceleration parameter $q_0$ and the density of
nonrelativistic matter $\Omega_{m0}$ reducing to just two of them in a flat
universe. 

The Stephani model under consideration is a simple generalization of a
flat Friedmann 
model and its parameter space can mimic (as far as the redshift-magnitude
relation is concerned) that of Friedmann with an important admission of the
effect of pressure gradient (acceleration) in the universe. 

The standard cosmological test -- a redshift-magnitude relation -- to second 
order in redshift $z$ for Friedmann models reads as (e.g. \cite{mar0}) 
\begin{eqnarray}
m_{\rm{B}} & = & M_{\rm{B}} + 25 
+ 5\log_{10}{cz} - 5\log_{10}{H_0}  \nonumber \\
& + & 1.086 \left( 1 - q_0 \right) z + 0.2715 \left[ 3 (1 + q_0)^2 - 4 (1 +
\Omega_{m0}) \right] z^2 + O(z^3)  ,
\end{eqnarray}
where $m_B$ is the apparent bolometric magnitude of a galaxy, $M_B$ is its
absolute magnitude and $z$ is the redshift. For Friedmann cosmologies the
following relations between the parameters are fulfilled 
\begin{eqnarray}
\Omega_\Lambda = \frac{\Lambda}{3H_0^2} = \frac{1}{2} \Omega_{m0} - q_0
,\hspace{1.cm} \frac{k}{H_0^2R_0^2} = \frac{3}{2} \Omega_{m0} - q_0 - 1  ,
\end{eqnarray}
where $\Omega_{\Lambda}$ is the density of cosmological constant $\Lambda$, 
$k$ the curvature index and $R_0$ the present value of the scale factor. The relation
(5) was tested by supernovae data \cite{42} and the best fit values of
the cosmological parameters in flat $(k = 0 = \Omega_{m0} + \Omega_{\Lambda} -
1 )$ universe were claimed to be 
\begin{eqnarray}
q_0 & = & - 0.55  ,\\
\Omega_{m0} & = & 0.3  ,\\
\Omega_{\Lambda} & = & 0.7  ,
\end{eqnarray}
for the Hubble'a constant 
\begin{equation}
H_0  =  63 \hspace{0.2cm} {\rm km s^{-1} Mpc^{-1}}  ,
\end{equation}
giving the best-fit age of the universe 
\begin{equation}
t_0  =  14.9 \hspace{0.2cm} {\rm Gyr}  .
\end{equation}.

The redshift-magnitude relation for Stephani universes has been found in
\cite{mar2}. Two exact cases were presented and the theoretical 
relations were plotted for a range of different parameter values. The
relations were obtained following the method of Kristian \& Sachs \cite{kris}
of expanding all relativistic quantities in power series and truncating at
a suitable order, though, one can use an exact relation too \cite{chris}. 
An analogous to (5) relation for the Stephani universe (1), to second
order in redshift $z$, reads as \cite{mar2,mar3}
\begin{eqnarray}
m_{\rm{B}} & = & M_{\rm{B}} + 25 + 5\log_{10}{cz} - 
5\log_{10}{\tilde{H}_0} + 1.086 \left( 1 - \tilde{q}_0 \right) z
\nonumber \\
& + & 0.2715 \left[ 3 (1 + \tilde{q}_0)^2 - 4 (1 +
\tilde{\Omega}_{m0}) \right] z^2 + O(z^3)  ,
\end{eqnarray}
where 
\begin{eqnarray}
\tilde{H}_0 & = & \frac{2 a\tau_{0} + 1}{a\tau_{0}^2 + \tau_{0}} ,\\
\tilde{q}_0 & = & - 4a \frac{a\tau_{0}^2 + \tau_{0}}{(2a \tau_0 + 1 )^2}  .
\end{eqnarray}

Equation (12) now takes the same functional form as equation (5),
as was similarly pointed out in \cite{mar2} with $\tilde{H}_0$ and 
$\tilde{q}_0$ replacing $H_0$ and $q_0$. We can think of $\tilde{H}_0$ 
and $\tilde{q}_0$ as a 
generalised Hubble parameter and deceleration parameter which are related to the
age of the universe in a different way from the Friedmann case. The key
question of interest here is therefore whether one can construct
generalised parameters, $\tilde{H}_0$ and $\tilde{q}_0$, which are in good
agreement with the supernovae data \cite{42} but which correspond to a value of $\tau_0$
which exceeds that Friedmann age with $H_0 = \tilde{H}_0$ and 
$q_0 = \tilde{q}_0$. More precisely, both relations (5) and (12) are equal
provided a generalized density parameter $\tilde{\Omega}_{m0} = (1/3) (1 +
\tilde{q}_0)$. Assuming the following values for the replaced parameters 
\begin{eqnarray}
\tilde{H}_0 & = & 63 \hspace{0.2cm} {\rm km s^{-1} Mpc^{-1}}  ,\\
\tilde{q}_0 & = & - 0.55  ,
\end{eqnarray}
we obtain the age of the universe in the Stephani model (1) to be 
\begin{equation}
\tau_0 = 18.67 \hspace{0.4cm} {\rm Gyr}   ,
\end{equation}
and $\tilde{\Omega}_{m0} = 0.15$.
Then we have an agreement with supernovae data, provided the non-uniform
pressure parameter is equal to 
\begin{equation}
{\rm a} = 12.3 \hspace{0.2cm} {\rm km^2 s^{-2} Mpc^{-1}}   ,
\end{equation}
which translates into the value of the acceleration scalar \cite{mar2} to be 
\begin{equation}
\dot{u} = -2 \frac{a}{c^2} r = - 2.73 \cdot 10^{-10} r {\rm Mpc}^{-1}  ,
\end{equation}
with $r$ being the radial coordinate of the model. Since the 
non-uniform pressure parameter $a$ is positive, then the high 
pressure region is at $r = 0$, while the low (negative) pressure regions are
outside the center, so the particles are {\it accelerated away from the
center}. This is similar effect as that caused by the positive cosmological
constant $\Lambda > 0$ in Friedmann models, although the physical mechanism
is somewhat different. 
However, in both cases it is worth to appeal to ideas from the theory of
elementary particles \cite{defects} and especially to the notion of the
energy of the vacuum. While in Friedmann models vacuum gives {\it constant
pressure} on every spatial section of constant time, in Stephani models 
it gives the pressure which {\it depends on spatial coordinates}. 

Finally, we emphasize that the result obtained here is based only on the
studies of the redshift-magnitude relation. The value of the non-uniform
pressure parameter in (18) should also be tested by the level of the
microwave background anisotropies and other astronomical data. 
  
Spherically symmetric models in which cosmic acceleration is also 
explained by the inhomogeneity in pressure (though except acceleration
admitting shear) have been considered recently by Pascual-S\'anchez
\cite{ps}. They form a larger class than Stephani models.

\begin{center}
{\bf Acknowledgments}
\end{center}

I acknowledge useful discussions with Richard Barrett, Chris Clarkson and
Martin Hendry.

\end{document}